\documentclass{aa}
\usepackage[dvips]{graphics}
\usepackage[dvips]{graphicx}

\newcommand{\teff}{T$_{\rm eff}$}

\newcommand{\nli}{$\log$~n(Li)}

\begin{document}

\thesaurus{1(10.15.2 NGC 3680; 10.15.2 IC 4651; 08.01.1; 08.09.3)}

\title{Evolution of lithium in solar--type stars: clues from intermediate age 
clusters
\thanks{Based on observations
carried out at the European Southern Observatory, La Silla, Chile}}

\author{S. Randich \inst{1}, L. Pasquini \inst{2}, and R. Pallavicini \inst{3}}

\institute
{Osservatorio Astrofisico di Arcetri, Largo Fermi 5, I-50125
Firenze, Italy\\e-mail: randich@arcetri.astro.it
\and European Southern Observatory, Karl-Schwartzschild Str. 2, D-85748,
Garching bei M\"unchen, Germany
\and Osservatorio Astronomico di Palermo, Piazza del Parlamento 1, I-90134
Palermo, Italy}

\offprints{S. Randich}

\date{Received / Accepted} 

\titlerunning{Evolution of lithium} 
\authorrunning{S. Randich et al.}

\maketitle
\begin{abstract}
We present Li abundances for 14 solar--type stars in the intermediate
age ($\sim 2$~Gyr) clusters IC~4651 and NGC~3680. The \nli~vs. effective
temperature distributions are compared with those of the similar age cluster
NGC~752, of the younger Hyades (600~Myr) and of 
the older M~67 (4.5~Gyr) and NGC~188 (6--7~Gyr) clusters.
Neither IC~4651 nor NGC~3680 show the dispersion in Li
which is observed in M~67. The 2~Gyr clusters have very similar
Li vs. \teff~distributions; in addition, stars in the upper envelope of the
M~67 distribution have the same Li content as stars in the 2~Gyr clusters,
suggesting that either they have not suffered any significant depletion between
$\sim$~2 and 4.5~Gyr or they had a much slower Li depletion. 
Mechanisms that lead to Li depletion on the main sequence are 
discussed in the light of these observations. None of the existing models
seem to reproduce well the observed features.


\end{abstract}

\section{Introduction}
Understanding the processes that lead to lithium destruction
on the main sequence (MS) provides
powerful diagnostics of stellar structure and evolution and  of
mixing mechanisms in stars.

\begin{center}
\begin{table*}
\caption{The sample}
\begin{tabular}{rccccl}\\ \hline
& & & & & \\
\multispan{6}{{\bf IC~4651}\hfill}\\
& & & & & \\
name & (B$-$V)$_0$ & EW(Li~{\sc i}) & \teff & \nli & RV \\
 (1)  &  &    (m\AA) &  (K)   &          & info. (2) \\          
  EG 7    &  0.56 & 70 $\pm$ 15  & 6061 & 2.7 $\pm 0.12$ & S \\
  EG 45   &  0.57 & 44 $\pm$ ~~7 & 6016 & 2.4 $\pm 0.12$ & S \\
  AT 38   &  0.54 & 35 $\pm$ ~~8 & 6122 & 2.4 $\pm 0.06$ &   \\
  AT 39   &  0.54 & 49 $\pm$ 10  & 6114 & 2.6 $\pm 0.06$ &   \\
  AT 1108 &  0.53 & 47 $\pm$ 10  & 6189 & 2.6 $\pm 0.06$ & SB \\
  AT 1109 &  0.62 & 61 $\pm$ 15  & 5814 & 2.4 $\pm 0.08$ &   \\
  AT 1225 &  0.51 & 64 $\pm$ ~~7 & 6235 & 2.8 $\pm 0.08$ & S \\
  AT 2207 &  0.59 & 44 $\pm$ 10  & 5908 & 2.3 $\pm 0.08$ &   \\
  AT 2105 &  0.54 & 48 $\pm$ 10  & 6110 & 2.5 $\pm 0.08$ & S \\
  AT 3226 &  0.58 & 49 $\pm$ ~~8 & 5967 & 2.4 $\pm 0.06$ & NM \\
  AT 4226 &  0.62 & 31 $\pm$ 10  & 5795 & 2.1 $\pm 0.08$ &  \\
& & & & & \\
\multispan{6}{{\bf(1)} From Eggen (\cite{eg71}) and Anthony-Twarog et al.
(\cite{at88}). Photometry comes from the latter source.\hfill}\\ 
\multispan{6}{{\bf(2)} 
Information on radial velocities (membership and binarity)
was kindly 
provided by Dr. B. Nordstr\"om. \hfill}\\
& & & & &\\
\multispan{6}{{\bf NGC~3680}\hfill}\\
& & & &  &\\
name & (B$-$V)$_0$ & EW(Li~{\sc i}) & \teff & \nli & RV \\
     &      &    (m\AA) & (K)   &              & info. \\          
 23   & 0.58 & 46$\pm$ 5 & 5971 & 2.41 $\pm$ 0.04 & SB1 \\
 60   & 0.59 & 48$\pm$ 6 & 5924 & 2.38 $\pm$ 0.05 & M? \\
 70   & 0.60 & 42$\pm$ 5 & 5884 & 2.28 $\pm$ 0.04 & M \\
 4114 & 0.58 & 51$\pm$ 6 & 5951 & 2.44 $\pm$ 0.05 & M? \\
& & & &  &\\
\multispan{6}{{\bf(1)} Star numbers, colors, and information on 
membership/binarity were taken from 
Nordstr\"om et al. (1997). \hfill}\\ \hline
\end{tabular}
\end{table*}
\end{center}

Surveys of Li among cluster and field stars
have shown that standard models (i.e., those that take into account
convecting mixing only) are in contradiction with many
of the observational features (see
Deliyannis \cite{del00}; Jeffries \cite{jef00}; Pasquini \cite{pas00}, 
and references therein for
recent reviews). Focusing on solar--type stars, at least two
results in strong disagreement with model predictions
have been found: {\it i)} these stars appear to deplete
Li while on the MS, in spite of the fact that 
their convective zones bases are too cool to burn Li;
{\it ii)} even more surprisingly, old,
otherwise similar solar--type stars show different amounts of Li depletion. 
Solar--type stars in the solar age, solar metallicity cluster M~67 (4.5~Gyr)
show a dispersion in Li abundances larger than a factor of 10
(Spite et al. \cite{spi87}; Garc\'\i a L\'opez et al. \cite{gar88}; 
Pasquini et al. \cite{pas97};
Jones et al. \cite{jon99}), in contrast with the tight relationship
between Li abundance and effective temperature (or mass) observed for 
single stars in the younger Hyades (age 600 Myr). 
The evolution from the Hyades to the age of the Sun
seems to be `bimodal', with a fraction ($\sim$ 60 \%) 
of stars depleting a relatively small amount of
lithium, and another fraction,
virtually similar to the other one as far as mass and chemical composition
are concerned, that undergoes a severe Li depletion. It is important to mention
that, whereas no Li data for other 
clusters coeval to M~67 are available, a similar
behavior is observed among old solar-type stars in the field 
(e.g., Pasquini et al. \cite{pas94}). The Sun with $\log$
n(Li) = 1.1 belongs to the class of Li-poor stars. In summary, not only we do
not understand the mechanism(s) that lead to Li depletion on the MS,
but these mechanism(s) seem to work
differentially for otherwise similar stars.

\begin{figure}
\resizebox{8.8cm}{!}{\includegraphics{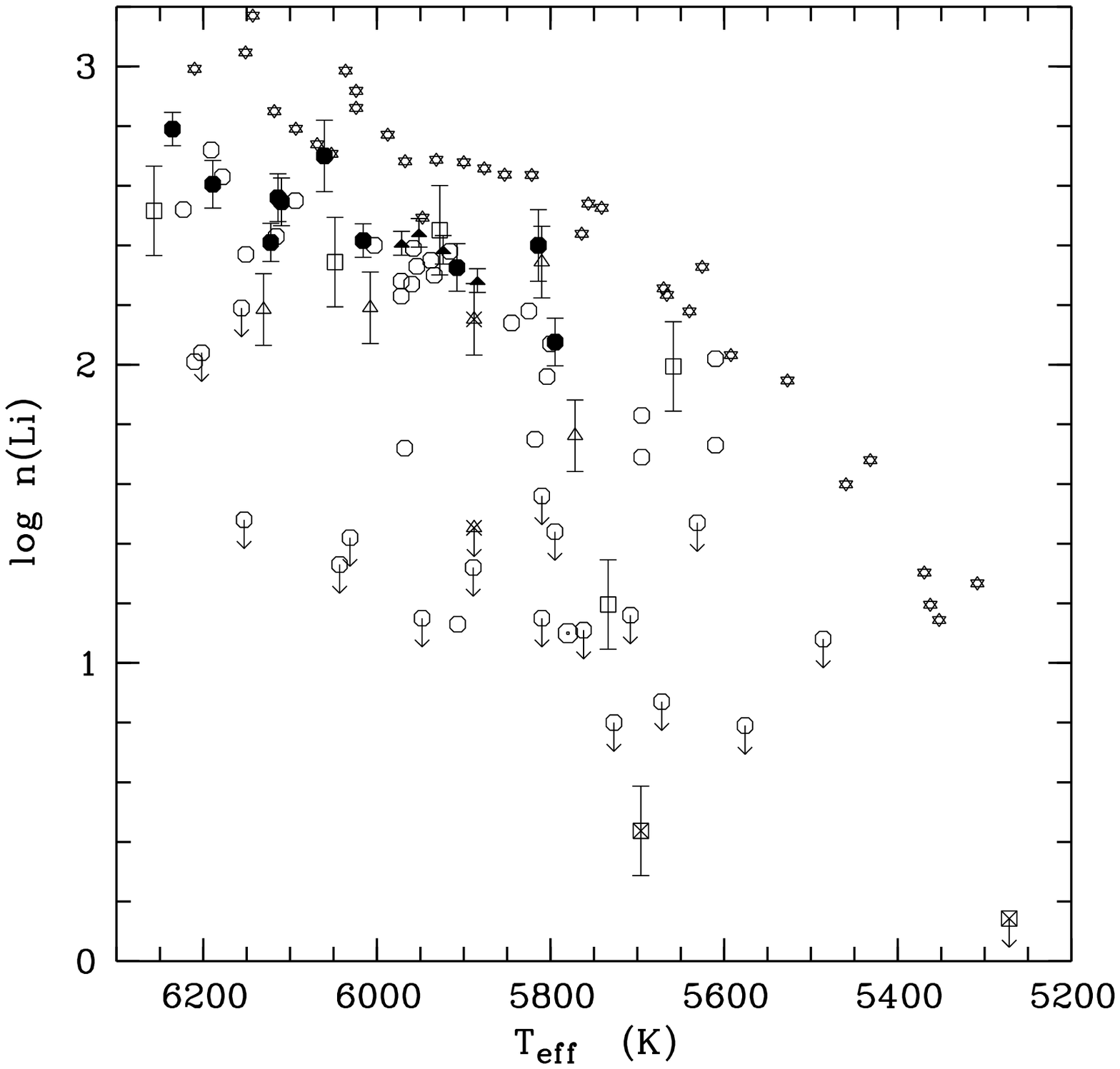}}
\vspace{-3.7cm}
\caption[]{\nli~vs. \teff~for our sample stars in IC~4651 and NGC~3680 (filled
circles and triangles, respectively), the Hyades (stars), M~67 (open circles),
NGC~752 (open squares), and NGC~188 (open triangles). 
M~67 data have been taken from Jones et al. (\cite{jon99}). The 
SB2 binary S1045 is not included in the figure.
Possible non--members in
NGC~752 and NGC~188 are indicated as crossed symbols. Errors in \nli~ take
into account only errors in EWs. The Sun is also shown in the figure.}
\end{figure}

\begin{figure}
\resizebox{8.8cm}{!}{\includegraphics{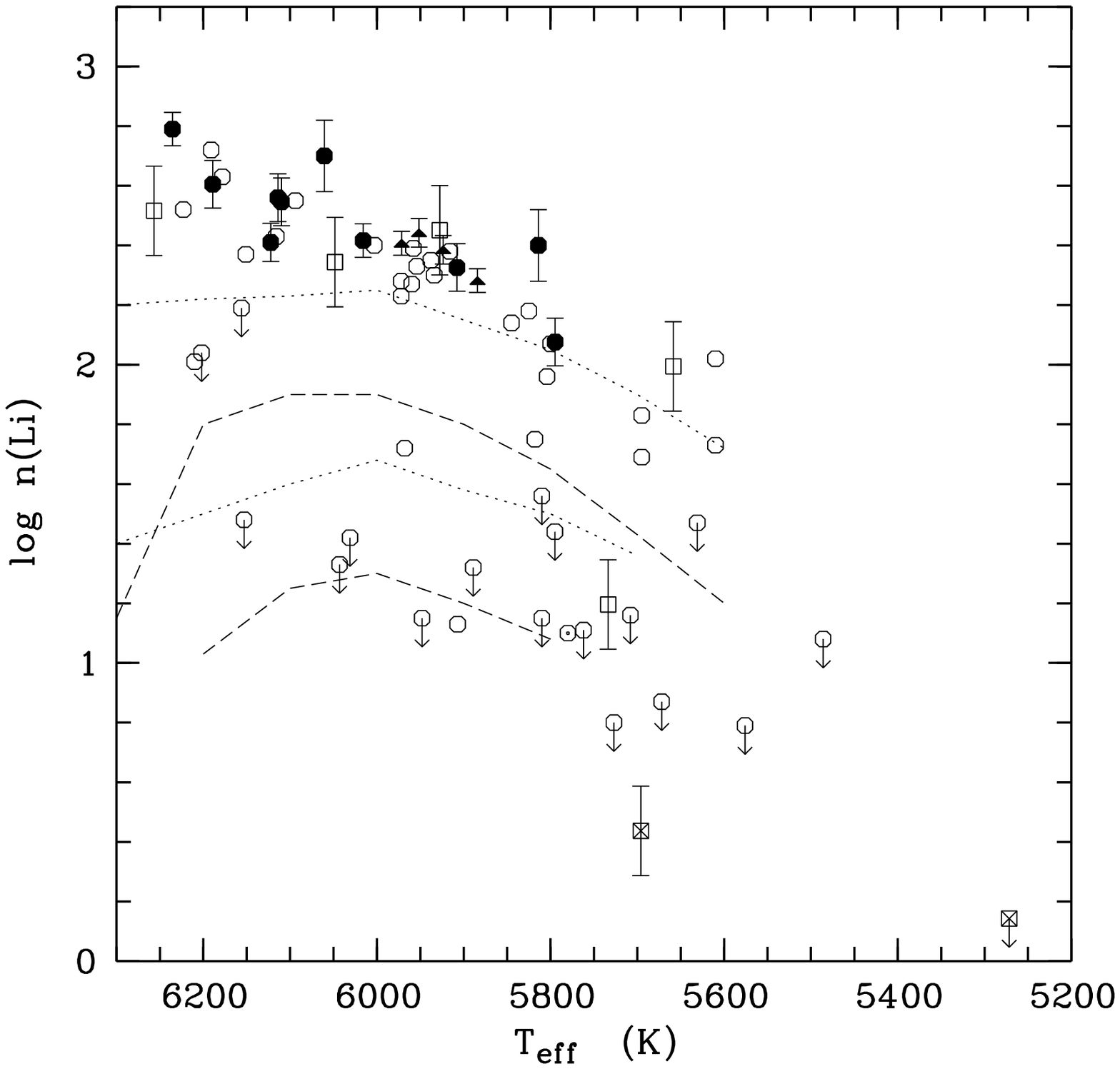}}
\vspace{-3.7cm}
\caption[]{\nli~vs. \teff~for IC~4651, NGC~3680, NGC~752, and M~67. Symbols are
the same as in Fig.~1.
Dotted and dashed curves represent the 1.7~Gyr and 4~Gyr isochrones for models
with rotationally induced mixing; upper
and lower curves at the same age
denote models with initial rotational velocities of 10
and 30 km/sec, respectively. 
}
\end{figure}
As a possible solution for the dispersion observed among M~67 stars,
Jones et al. (\cite{jon99}) speculated that the mixing 
is driven by MS spin--down and angular momentum loss; in this 
case, the dispersion in Li would be due to different initial 
rotation rates. If this were the case, a spread in Li should
be seen already at ages of 2 Gyr or younger 
(e.g., Pinsonneault \cite{pin97}).
More in general, in order to explain the observational evidences, several
models have been developed in the last decade, involving more complex physics
or/and additional mixing processes; the validation of these models
still needs strong observational support, in particular at ages
intermediate between the Hyades and M~67 where very few data are available.
Determining Li in a significant sample of
stars in intermediate age clusters is thus extremely
important: so far Li measurements for clusters of this age
have been obtained only for very few 
objects in NGC~752 (Hobbs \& Pilachowski, \cite{hp86}).

We present here Li abundances for a sample of solar--type
stars in the open clusters IC~4651 and NGC~3680.
These clusters, for which metallicities close to solar have
been derived (Friel \cite{fr95}),  have estimated ages 
of 1.5 -- 2 Gyr (e.g., Meynet et al. 
\cite {mey93}; Friel \cite{fr95}); these data,
therefore, allow us to address the problem of Li depletion from the age
of the Hyades to that of M~67
and to put observational constraints 
on the mechanisms responsible for MS Li depletion and on their timescales.
\section{Observations and analysis}
The observations of both clusters were carried out
at ESO, La Silla.  
Most of IC~4651 stars were observed during three observing
campaigns in June 1997,
June 1998, and May 1999. The 3.6~m telescope with the CASPEC spectrograph
was used; the standard grating (31.6 lines/mm), 
the red cross-disperser, the long camera, and ESO CCD \#37 
(1024 $\times$ 1024 24$\mu$ pixels) were employed. 
%were 200, 180, and 280~$\mu$ (1.4, 1.3, and 2 arcsec on the sky), resulting in
%resolving powers $R \sim$ 29,000, 31,000, 20,000, respectively. The use of
%different slit widths was due to different seeing conditions.
A few additional IC~4651 spectra were taken in sparse observing runs
with FEROS at the 1.5~m ESO telescope and with CASPEC.
The spectra of NGC~3680 were obtained using CASPEC.
Different slit widths (from 1.4 to 2 arcsec)
were used in the various CASPEC runs due to
different seeing conditions. Correspondingly, we obtained spectra with
resolving powers ranging from $R \sim 20,000$ to $R \sim 30,000$.
The FEROS spectra have resolution $R=48,000$.
The S/N of the spectra is in the range $\sim$ 30 to 60.

Data reduction for the CASPEC spectra was performed using the
{\sc echelle} context within MIDAS and following the usual steps.
FEROS spectra were reduced using the FEROS data reduction pipeline.

Lithium abundances were obtained in the same fashion as in Jones et al. 
(\cite{jon99})
for M~67. Namely, effective temperatures were derived from dereddened
B$-$V colors using the relationship of Soderblom et al. (\cite{sod93}).
Reddening values E(B--V)$=0.05$ and $0.083$ were assumed for NGC~3680 and
IC~4651, respectively.
Abundances were derived
from measured equivalent widths (EWs) using the curves of growth (COGs)
of Soderblom et al. (\cite{sod93}). When the Li~{\sc i} 6707.81
doublet was blended with the Fe~{\sc i}~$\lambda$~6707.44\\line, we estimated
the contribution of the latter using the prescription of Soderblom et al.
(\cite{sod93}). 
Information on the sample stars together with our Li measurements are
given in Table~1.
We carried out a similar analysis for NGC~752 and NGC~188, using the
published Li EWs (from Hobbs \& Pilachowski \cite{hp86}, \cite{hp88}) 
and the most updated sources for photometry and membership. Hyades data from
Thorburn et al. (\cite{tho93}) were also re-analyzed in a consistent way.
We stress therefore
that the Li abundances used in this 
letter are all on the same abundance scale.

Besides dwarf stars, we observed giants and turn--off
stars in the two clusters; these data will be used to address
different issues and they will be presented in a separate paper.
\section{Results}
In Figure~1 we plot \nli~vs. effective temperature
for IC~4651 and NGC~3680
together with the Hyades, NGC~752, M~67, and NGC~188.
Four major features are evident in the figure: {\bf 1.} 
The three about coeval clusters,
IC~4651, NGC~3680, and NGC~752, have very similar Li vs. \teff~ 
distributions, with only a couple of IC~4651 stars lying somewhat
above the average trend (but within 1$\sigma$ error bar from it).
Their average Li at a given \teff~is about 0.4~dex
(or a factor 2.5 in abundance) below the Hyades; {\bf 2.} 
if we consider the upper envelope of the Li vs. \teff~patterns, it is
evident that the less Li depleted stars in M~67 have very similar
abundances as the stars in the $\sim 2$~Gyr clusters. 
The NGC~188 (age 6--7~Gyr) 
sample is too small and the error bars too large to allow definite
statements, but there might be an additional slight Li decrease with respect
to M~67;
{\bf 3.} {\it none} of the three
intermediate age clusters shows any significant dispersion for
stars warmer than
$\sim$ 5800~K; there are no stars cooler than this in NGC~3680 and IC~4651,
while the three NGC~752 stars cooler than 5800~K have significantly
different Li abundances (but one is possibly a non--member); {\bf 4.} as
now well known, a large fraction of
stars in M~67 have suffered much more Li depletion
than stars defining the upper envelope of this cluster.

Under the assumption that all the clusters had the same initial Li abundance,
the four points listed above in turn suggest that: {\bf i.}
Li depletion among solar--type stars
older than the Hyades is a single function of age
up to $\sim$ 1.5--2 Gyr and \teff~$\sim$~5800~K;
{\bf ii.} the mechanism
which drives Li depletion between the Hyades and the 2~Gyr old clusters
appears to saturate, or to considerably slow down, at ages of
2~Gyr. If stars continued to deplete Li at
the same rate as between the Hyades and the 2~Gyr clusters,
one would expect a maximum Li abundance 
(in the range $\sim 6100 <$\teff$<5800$~K) of the order of 
\nli$\sim 2.1$ at the M~67 age and of the order of $\sim 1.95$
at the NGC~188 age. Both clusters show a higher maximum Li abundance; 
the alternative
hypothesis can be made however that Li-rich M~67 
stars suffered a slower overall
Li destruction than inferred from younger clusters; {\bf iii.} 
Li depletion appears to stop or to slow down
only for a fraction of the stars, while the other 
fraction suffers significant additional depletion. 
Additional depletion may start at ages younger than 2~Gyr for stars cooler than 5800~K. On the other hand, we cannot exclude that the Sun 
itself at 2~Gyr still had a Li
abundance a factor of 10 higher than its present value.

Before trying to interpret these results,
we must check whether and to which extent the lack of a major dispersion among
the IC~4651 and NGC~3680 clusters
may be due to low number statistics. 37 M~67 members are included
in the 6200 $\geq$~\teff~$\geq 5800$~K range: excluding the SB2 binary S1045,
11 of them ($\sim$ 30 \%) are Li--poor,
23 ($\sim$ 62 \%) are Li-rich, and one has a 
very high Li upper limit. The IC~4651 $+$ NGC~3680 merged sample in the
same temperature range contains 14 stars; more specifically,
our sample contains about
half of the total population of solar--type stars in IC~4651 and all the
presently known solar--type stars in NGC~3680.
In order to have a similar fraction
of Li-poor/Li rich stars as in M~67
one would have to assume that all the Li-rich stars fell
in our (unbiased) sample or, viceversa, that most of  
the stars that were not observed by us are Li-poor. Whereas this possibility
cannot be entirely excluded, it would represent a very unusual coincidence.

\section{Constraints on MS Li depletion mechanism(s)}

Basically four possible processes have been suggested in the literature
for Li depletion on the MS: namely, microscopic diffusion, mass
loss, and slow mixing either induced by rotation 
or driven by gravity waves.

{\bf a)} Diffusion is a slow mechanism and
could in principle explain the smooth decay of the maximum Li abundance
between the Hyades and the 2~Gyr clusters,
as well as the lack of a star-to-star scatter
up to at least 2~Gyr. Diffusion has been shown to be able to explain Li in
the Sun, where He diffusion is validated by heliosismology (Vauclair 
\cite{vau98}). However, diffusion is probably far too slow:
model calculations suggest that
due to the increasing
depth of the convective zone, below 6400~K the diffusion timescales are
very large, comparable to the evolutionary timescale, and
thus no significant Li depletion should occur before the star reaches the very
end of its MS lifetime (Michaud \cite{mic86}). 
In addition, it would be difficult to explain
the ``saturation" of Li depletion that seems to occur 
for clusters older than 2~Gyr.
{\bf b)} Mass loss has been shown not to be able to reproduce the Hyades Li vs.
\teff~ distribution (Swenson \& Faulkner \cite{swe92}) 
and, according to the same
authors, it  would be even more difficult to reproduce the NGC~752 pattern;
by extension, it is reasonable to
expect that diffusion would not be able to account for the IC~4651 and NGC~3680
Li patterns. In any case, contrary to diffusion, mass loss is probably
a too fast mechanism to account for the slow decrease of the maximum Li 
abundance.
{\bf c)} Both rotation (due to angular momentum loss)
and waves driven mixing mechanisms are rather slow processes
and, in this sense, have the right timescales to
explain the decay between the Hyades and the 
2~Gyr clusters. {\bf c1)} Montalb\'an \& Schatzman (\cite{mon96})
were able to reproduce
the Hyades pattern at 600~Myr with somewhat
simplified models including gravity waves and
a certain amount of PMS depletion; the speculation therefore
can be made that the distribution of older clusters may be fitted
as well. However, it is not clear whether
this mechanism would saturate
or not; {\bf c2)} as to rotational mixing,
quoting from Pinsonneault (\cite{pin97}), ``Theoretical models of rotational
mixing in solar analogs produce a rate of depletion that decreases
with age as stars spin down". Thus, it is plausible that Li destruction
eventually stops (or becomes very slow) when stars are completely spun-down.
However, whereas ``The existence of a dispersion is a prediction of rotational
mixing",
our results indicate that a significant dispersion develops only very
late during MS evolution. In Figure~2 we show again \nli~vs. \teff~
for IC~4651, NGC~3680, NGC~752, and M~67. 
The rotationally induced mixing isochrones at 1.7 and 4~Gyr of 
Deliyannis \& Pinsonneault (\cite{del97};
see also Stephens et al. \cite{ste97}) for two initial rotational
velocities are also shown in the plot.
The figure indicates that the virtually
null Li destruction for Li-rich stars observed
between the 2~Gyr clusters and M~67
is not predicted by the models and, more in general, that the upper
envelopes of the observed distributions are not well reproduced by
the isochrones corresponding to initial velocities of 10~km/sec,
too much Li depletion being predicted by the latter models.
In addition, as the figure shows, the models predict that a large 
difference in Li  between stars with initial velocities of 10 and 30~km/sec
should be present already at ages of $\sim$
1.7~Gyr. On the other hand, the 4~Gyr model with an initial rotational
velocity of 30 km/sec fairly well reproduces the lower envelope of M~67.

In summary, both diffusion and
mass loss can probably be excluded as processes affecting in a major
way Li depletion on the MS. If any, diffusion may have some effect
between M~67 and NGC~188, if the maximum Li abundance in the latter
cluster is really slightly below that of M~67.
Gravity waves are a viable mechanism, but 
they cannot reproduce
the spread observed in M~67; more work needs to
be carried out on model calculations to check whether they could 
fit the intermediate age cluster patterns. Rotational mixing has in principle
the requirements to meet the observational constraints,
but currently 
available models do not fit quantitatively the observational data.
In order to get a better agreement between observations and models, the
latter ones should predict an even slower Li depletion for stars with
low initial velocities (of the order of 10 km/sec --models with even lower
initial velocities should also be computed)
and, furthermore, they should predict no (or very small) additional depletion
after 2~Gyr for these stars 
(it is conceivable that stars
starting their MS lifetimes with low rotational velocities should have
a small degree of internal differential rotation by that age). In order to
reproduce the lack of a spread at 2~Gyr, stars that
start their MS lifetimes as rapid rotators should deplete
the same amount of Li as initially slow rotators up to $\sim$ 2~Gyr; afterwards
rotation driven mixing should become much faster.

The above discussion is based on the assumption that the three middle--aged
clusters are representative of stars at 2~Gyr, while M~67 is representative
of stars at the solar age. In other words, we gave for granted that the
M~67 and the middle--aged clusters had similar initial properties (in
particular a similar distribution of rotation rates) 
and that M~67 at 2~Gyr had a similar Li distribution as
the intermediate clusters or, viceversa, that the 
latter clusters would show at the M~67 age a Li pattern similar
to that of M~67. However, we cannot exclude that M~67 and the intermediate
age clusters were characterized by different initial conditions
and, consequently, underwent a different rotational history and Li evolution.
For example, the intermediate age clusters may have had a very small initial
rotational spread with most of the stars being very slow rotators
(but available rotational data for young clusters do not 
support this ideas); or, as originally suggested by Garc\'\i a L\'opez
et al. (\cite{gar88}), there may have been more than one burst of
star formation within M67.
Obviously, additional intermediate age and
old clusters should be observed in order to test
these hypotheses. 

\begin{acknowledgements}
We are grateful to Drs. J.~Andersen and B.~Nordstr\"om 
for providing information
on radial velocities and membership for IC~4651 prior to publication.
We thank the referee, Dr. F. Spite, for useful comments on the manuscript.
\end{acknowledgements}

{}

\end{document}